# The Effect of Warm-Glow on User Behavioral Intention to Adopt Technology: Extending the UTAUT2 Model

Antonios Saravanos[1(✉)], Neil Stott[2], Dongnanzi Zheng[1], and Stavros Zervoudakis[1]

[1] New York University, New York, NY 10003, USA
{saravanos,dz40,zervoudakis}@nyu.edu
[2] Cambridge Judge Business School, Cambridge CB2 1AG, UK
n.stott@jbs.cam.ac.uk

**Abstract.** In this study, we enhance the Unified Theory of Acceptance and Use of Technology (UTAUT2) by incorporating the warm-glow phenomenon to clarify its impact on user decisions regarding the adoption of technology. We introduce two additional constructs aimed at capturing both the external and internal aspects of warm-glow, thus creating what we refer to as the UTAUT2 + WG model. To evaluate the effectiveness of our model, we conducted an experimental study in which participants were presented with a scenario describing a hypothetical technology designed to evoke warm-glow sensations. Using the partial least squares method, we analyzed the collected data to assess our expanded model. Our findings indicate that warm-glow significantly influences user behavior, with the internal aspect having the strongest influence, followed by hedonic motivation, performance expectancy, and finally the external aspect of warm-glow. We conclude by discussing the implications of our research, acknowledging its limitations, and suggesting directions for future exploration.

## 1 Introduction

Understanding the factors that influence a user's intention to adopt a technology and their relative importance is crucial for organizations seeking to refine their strategies. Technology acceptance modeling provides a well-established framework for gaining such insights, with the extended Unified Theory of Acceptance and Use of Technology (UTAUT2) representing the latest version [32]. This model has been customized to accommodate different contexts (e.g., consumer versus workplace), cultural differences (e.g., East versus West), and technological applications (e.g., mobile banking, biometric authentication). Numerous systematic studies have documented the model's adaptability and application across various domains (e.g., Tamilmani et al. [30]). However, one area that has not been extensively explored within the UTAUT2 framework is the adoption of technologies that evoke a warm-glow sensation among potential users.

Warm-glow refers to the positive feeling one experiences when doing something good for others, regardless of whether the motivation behind the action is selfish or altruistic. This sentiment is categorized into two forms: extrinsic warm-glow (EWG) and intrinsic warm-glow (IWG) (see Shakeri and Kugathasan [28]). While the phenomenon of warm-glow has been explored in relation to technology adoption through frameworks such as Ajzen's [1] Theory of Planned Behavior (e.g., Gamel et al. [10]) and the Technology Acceptance Model (TAM) (e.g., Saravanos et al. [25]), it is important to note distinctions between them. The Theory of Planned Behavior is a general model for understanding user adoption behavior, not specifically tailored for technology. On the other hand, the TAM is customized for technology use but is not the most current model available. In contrast, the UTAUT2 model represents an evolution of various technology adoption models, including TAM, and is considered the preferred framework for contemporary technology adoption research.

In this study, we expand the UTAUT2 model to include the concept of warm-glow. We introduce two factors, namely perceived extrinsic warm-glow (PEWG) and perceived intrinsic warm-glow (PIWG), which are adapted from the research of Saravanos et al. [26]. These factors aim to capture the dual aspects of warm-glow: extrinsic and intrinsic. We refer to this enhanced model as UTAUT2 + WG. Subsequently, we utilize an experimental approach to evaluate the effectiveness of the model and examine the influence of warm-glow on users' behavioral intentions to adopt a technology.

## 2 Materials and Methods

Given that technology acceptance modeling is primarily confirmatory, the first step involves establishing a model, specifically the UTAUT2 + WG model, as outlined in this section. Subsequently, we provide details of the experiment conducted to clarify the role of warm-glow in the context of technology adoption.

### 2.1 Development of the Hypotheses and Model

In this subsection, we will review the constructs included in our proposed UTAUT2 + WG model, as shown in Fig. 1. Since our model is based on the UTAUT2 framework [32], we initially introduce and integrate the seven constructs from that framework: effort expectancy (EfEx), facilitating conditions (FacCon), habit (Hab), hedonic motivation (HeMo), performance expectancy (PerEx), price value (PriVal), and social influence (SocIn). These constructs have been identified as positively influencing a user's behavioral intention (BehInt) to adopt a technology, a concept we also integrate into our model (refer to Venkatesh et al. [32]). The first of these constructs, performance expectancy, as defined by Venkatesh et al. [32], refers to "the extent to which using a technology will provide benefits to consumers in performing certain activities". The second construct, effort expectancy (EfEx), is defined as "the degree of ease associated with consumers' use of technology" [32]. The third factor, social influence (SocIn), is presented as "the extent to which consumers perceive that important others (e.g., family and friends) believe they should use a particular technology" [32]. Facilitating conditions (FacCon), the fourth construct, is defined as "consumers' perceptions of the resources and support available to perform a behavior". The fifth construct is hedonic motivation (HeMo), which "is defined as the fun or pleasure derived from using a technology" [32]. Price value (PriVal) is defined as "consumers' cognitive trade-off between the perceived benefits of the applications and the cost for using them" [32]. Lastly, habit (Hab) is presented as "a perceptual construct that reflects the results of prior experiences" [32]. The model also includes three moderators: age, experience, and gender.

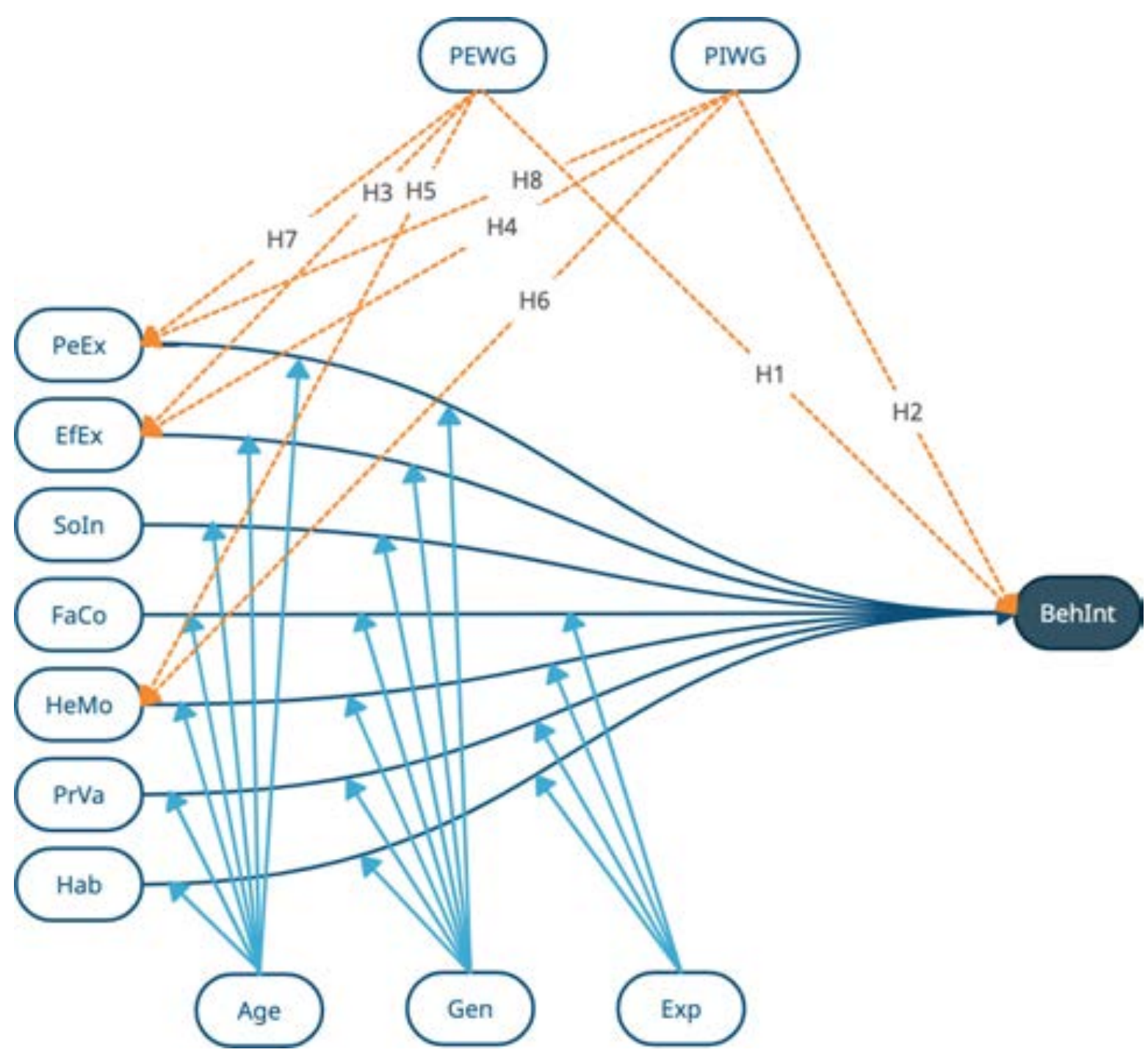

**Fig. 1.** Illustration of our proposed model.

### Incorporating Warm-Glow into Our Model

We complement these constructs by adding two additional dimensions, PEWG and PIWG, to capture the different aspects of warm-glow as perceived by end-users. Extensive research has shown the impact of warm-glow on consumer decision-making. Some studies, particularly in marketing literature, examine how cause-related marketing triggers warm-glow (e.g., Chaabane and Parguel [22]). Others explore the intrinsic dimension of warm-glow by using the Theory of Planned Behavior as a general adoption model (e.g., Azalia et al. [3], Bhutto et al. [4], Gamel et al. [10], Ma and Burton [21], and Sun et al. [29]). Some studies specifically focus on the extrinsic dimension (e.g., Griskevicius et al. [11]). Notably, Saravanos et al. [25] in 2022 investigated both dimensions of warm-glow using the Technology Acceptance Model 3 (TAM3). They found that both dimensions significantly influence user behavioral intention to adopt technology. Specifically, within the TAM3 context, the perception of IWG holds the second greatest influence following the technology's perceived usefulness, while the perception of EWG holds the fourth greatest influence, preceded by subjective norm. These studies provide ample rationale for integrating such factors into our model. Consequently, we propose the following hypotheses: **H1:** *PEWG positively influences BehInt*; **H2:** *PIWG positively influences BehInt*.

**The Effect of Warm-Glow on Perceived Usability**
Previous studies examining the influence of warm-glow on factors related to the perceived usability of a technology, such as EfEx, HeMo, and PeEx, are scarce. However, we found one study by Saravanos et al. [25] that implicitly investigated how users' perception of warm-glow affects their perception of the usefulness and ease of use of a technology. They discovered that the perception of EWG led to a perception of greater usefulness of the technology. Additionally, while not directly focusing on warm-glow, other research has shown that various factors can impact the perception of usefulness, ease of use, and hedonic motivation. For instance, Rodrigues et al. [24] found in their study that socialness influences perceived usefulness, even though their primary focus was on whether ease of use contributes to the perception of enjoyment. In this context, and for comprehensiveness, we propose the following six hypotheses: **H3:** *PEWG positively influences EfEx*; **H4:** *PIWG positively influences EfEx*; **H5:** *PEWG positively influences HeMo*; **H6:** *PIWG positively influences HeMo*; **H7:** *PEWG positively influences PeEx*; **H8:** *PIWG positively influences PeEx*.

**Confirming the Uniqueness of the PEWG and PIWG Factors**
It is important to evaluate whether the warm-glow factors we are incorporating into our model duplicate the functions of existing constructs in the UTAUT2 model. We have identified two constructs that seem somewhat similar and require further investigation. Specifically, we have examined whether PEWG could serve as a substitute for SoIn, as both constructs reflect the pleasure derived from others recognizing one's actions. However, SoIn addresses the overall situation while PEWG focuses specifically on the extrinsic aspect of warm-glow. Additionally, we have explored whether HeMo could replace PIWG, as both factors measure the pleasure derived from an action. It is worth noting that HeMo relates to the general situation while PIWG specifically targets the intrinsic aspect of warm-glow. For comprehensive analysis, we therefore propose the following two hypotheses: **H7:** *PEWG is a substitute to SoIn with respect to BehInt;* **H8:** *PIWG is a substitute to HeMo with respect to BehInt.*

## 2.2 Data Collection, Sample Characteristics, and Instrument

We used the Qualtrics platform to conduct a web-based experiment to test the proposed model. A total of 279 participants successfully completed the experiment. In terms of gender, males made up the majority at 59.5% (166 individuals), while females accounted for 40.5% of the sample, with 113 individuals. In terms of age, the majority of participants were between 26–45 years old. The highest proportion (30.82%) were aged 36–45 years, followed by 31–35 years (16.13%) and 26–30 years (14.7%). The age group least represented in the sample was 18–25 years, making up 8.24% of the participants. Regarding salary, the majority of participants earned between \$30,000 and \$49,999 annually (24.01%). Subsequently, 20.07% earned between \$50,000 and \$69,999, and equal proportions (15.05%) earned between \$70,000 and \$89,999 or \$90,000 or more.

The least common salary bracket was less than $10,000, at 6.09%. In terms of education, nearly half of the sample held a bachelor's degree (46.59%). The next largest group (20.79%) attended some college but did not earn a degree. Smaller groups included those with a professional degree (e.g., MBA, MFA, JD, MD) at 1.43% or a doctoral degree (e.g., PhD, EdD, DBA) at 2.15%. According to Chin and Newsted [6], our sample size is considered adequate for use with such models as it exceeds 150, which they consider to be large. Each participant provided consent before participating. Afterward, they were presented with a vignette (adapted from Saravanos et al. [26]) that described a hypothetical technology product designed to evoke a warm-glow feeling. Participants then completed a survey that included rating scales adapted from Venkatesh et al.'s [32] UTAUT2 instrument, in addition to demographic questions. Warm-glow questions proposed by Saravanos et al. [26] were also included to measure the two dimensions of warm-glow. Except for demographic questions, responses were collected on a 7-point Likert scale, ranging from strongly disagree to strongly agree.

## 3 Analysis and Results

To analyze the collected data, we employed Structural Equation Modeling (SEM), a method that extends traditional linear modeling techniques [7]. SEM is esteemed for its capacity to handle intricate models and is characterized as "used to explain multiple statistical relationships simultaneously through visualization and model validation" [7]. There are two primary variants of SEM: Covariance Based (CB) and Partial Least Squares (PLS). Hair et al. [13] provide guidance on when each should be utilized, suggesting that "PLS-SEM" is suitable for studies that are exploratory or an extension of existing structural theory. Since our goal was to extend an existing theory (UTAUT2 with warm-glow), we chose to use PLS-SEM for our analysis. Other factors that influenced this decision were PLS-SEM's effectiveness with "complex models", "non-normal data distributions", and "small sample sizes" [2]. PLS-SEM relies on two components: a measurement model that "specifies the relations between a construct and its observed indicators", and a structural model that "specifies the relationships between the constructs" [14].

The use of PLS-SEM consists of two consecutive phases, as described by Sarstedt et al. [27]. The initial phase involves assessing a pre-existing measurement model. Once the measurement model is confirmed, showing the reliability and validity of the measures, the second phase begins, where the corresponding structural model is tested. We provide further details on both phases in the following subsections.

### 3.1 Measurement Model

To evaluate our measurement model, we first assessed convergent validity, followed by construct reliability. For convergent validity, we examined factor loadings to determine the degree of correlation, as suggested by Sarstedt et al. [27], and removed any manifest variables with a loading value below 0.7 from our model. Convergent validity measures the extent to which two or more variables are correlated, indicating their theoretical association. Factor loadings indicate the level of correlation between an observed variable (manifest) and the corresponding latent variable (construct). Initially, we identified two factor loadings below the threshold: Hab4 with a loading of 0.586 and FaCo4 with a loading of 0.697. In such cases, the recommended course of action is to eliminate those factor loadings, which we did, before reassessing convergent validity. All the remaining factor loadings were at or above 0.7. Additionally, we examined the associated t-statistics, which measure the statistical significance of the factor loadings. All t-statistics were significant at the 0.01 level ($p < 0.01$), indicating that the factor loadings were statistically significant and not due to chance. Furthermore, we calculated the Average Variance Extracted (AVE), which reflects the average variance explained by a construct's manifest variables [9]. An AVE value of 0.5 or higher denotes an acceptable level of convergent validity [14]. All AVE values exceeded this threshold. In summary, all constructs demonstrated satisfactory convergent validity, evidenced by their high factor loadings, significant t-statistics, and AVE values, affirming the suitability of each item as a measure of its underlying construct and the validity of the scale used in this research. Subsequently, we proceeded to test construct reliability for each factor, relying on Cronbach's Alpha and Composite Reliability (CR) coefficients. All factors exhibited good to excellent reliability, with Cronbach's Alpha (ranging from 0.717 to 0.978) and CR values (ranging from 0.841 to 0.986) exceeding the widely accepted threshold of 0.7 [13]. Such high reliability scores suggest that the measures used are appropriate and consistent. Detailed values can be found in Table 1.

**Table 1.** Summary of Loadings, Convergent Validity and Construct Reliability Testing.

| Factor | Item | Loading | t-Statistic | AVE | Cronbach's Alpha | CR |
|---|---|---|---|---|---|---|
| BehInt | BehInt1 | 0.977 | 233.217* | 0.958 | 0.978 | 0.986 |
| | BehInt2 | 0.974 | 228.204* | | | |
| | BehInt3 | 0.985 | 428.811* | | | |
| EfEx | EfEx1 | 0.914 | 50.122* | 0.790 | 0.912 | 0.938 |
| | EfEx2 | 0.894 | 34.539* | | | |
| | EfEx3 | 0.906 | 51.658* | | | |
| | EfEx4 | 0.841 | 14.799* | | | |
| FaCo | FaCo1 | 0.898 | 31.757* | 0.793 | 0.871 | 0.920 |
| | FaCo2 | 0.871 | 27.038* | | | |
| | FaCo3 | 0.902 | 46.179* | | | |
| | HeMo1 | 0.966 | 193.444* | 0.906 | 0.948 | 0.967 |
| HeMo | HeMo2 | 0.951 | 118.772* | | | |
| | HeMo3 | 0.939 | 66.206* | | | |
| Hab | Hab1 | 0.739 | 13.706* | 0.637 | 0.717 | 0.840 |
| | Hab2 | 0.831 | 23.345* | | | |
| | Hab3 | 0.821 | 22.925* | | | |

**Table 1.** (*continued*)

| Factor | Item | Loading | t-Statistic | AVE | Cronbach's Alpha | CR |
|---|---|---|---|---|---|---|
| PeEx | PeEx1 | 0.833 | 38.489* | 0.805 | 0.919 | 0.943 |
| | PeEx2 | 0.900 | 54.180* | | | |
| | PeEx3 | 0.927 | 90.445* | | | |
| | PeEx4 | 0.925 | 76.399* | | | |
| PEWG | PEWG1 | 0.929 | 109.028* | 0.855 | 0.915 | 0.947 |
| | PEWG2 | 0.922 | 82.370* | | | |
| | PEWG3 | 0.923 | 80.025* | | | |
| PIWG | PIWG1 | 0.927 | 75.986* | 0.847 | 0.910 | 0.943 |
| | PIWG2 | 0.903 | 42.439* | | | |
| | PIWG3 | 0.932 | 68.756* | | | |
| PrVa | PrVa1 | 0.792 | 7.597* | 0.834 | 0.913 | 0.937 |
| | PrVa2 | 0.966 | 38.690* | | | |
| | PrVa3 | 0.971 | 38.079* | | | |
| SoIn | SoIn1 | 0.976 | 223.845* | 0.941 | 0.969 | 0.980 |
| | SoIn2 | 0.958 | 90.195* | | | |
| | SoIn3 | 0.977 | 220.751* | | | |

*$p < 0.01$.

### 3.2 Structural Model

Given a satisfactory measurement model, we proceeded confidently to evaluate the structural model. Regarding the variable of behavioral intention, our model yielded an $R^2$ of 0.706, explaining 70.6% of the variance, with an adjusted $R^2$ of 0.660. Interpretation of $R^2$ values for endogenous latent variables in the structural model can follow two guidelines. Chin [5] describes an $R^2$ greater than or equal to 0.19 and less than 0.33 as weak, an $R^2$ greater than or equal to 0.33 and less than 0.67 as moderate, and an $R^2$ greater than or equal to 0.67 as substantial. On the other hand, Hair et al. [13] suggest an $R^2$ greater than or equal to 0.25 and less than 0.50 as weak, an $R^2$ greater than or equal to 0.50 and less than 0.75 as moderate, and an $R^2$ greater than or equal to 0.75 as substantial. In this instance, following Chin's [5] guidelines, our $R^2$ value for BI can be described as substantial, and following Hair et al.'s [13] guidelines as moderate. Regarding the direct effects, the antecedents of the BI factor that are statistically significant, in order of decreasing strength, are as follows: PIWG ($\beta = 0.359$; $p < 0.01$), HeMo ($\beta = 0.329$; $p < 0.01$), and PeEx ($\beta = 0.302$; $p < 0.01$). When considering the total effects, the antecedents of the behavioral intention factor that are statistically significant, in order of decreasing strength, are PIWG ($\beta = 0.491$; $p < 0.01$), HeMo ($\beta = 0.329$; $p < 0.01$), PeEx ($\beta = 0.302$; $p < 0.01$), and PEWG ($\beta = 0.278$; $p < 0.01$). Therefore, a 1-unit increase in PEWG resulted in a 0.278-unit increase in BehInt, which is consistent with H1. Similarly, each 1-unit increase in PIWG resulted in a 0.491-unit increase in BehInt, which is consistent with H2.

In terms of the variable of effort expectancy, our model produced an $R^2$ of 0.196, explaining 19.6% of the variation, with an adjusted-$R^2$ of 0.190, which, according to Chin's guidelines [5], falls into the weak category. Only one construct, PIWG, was found to be statistically significant, and both direct effects and total effects had identical values ($\beta = 0.418$; $p < 0.01$). Therefore, an increase of 1 unit in PIWG corresponded to a 0.418-unit increase in effort expectancy, supporting H4. However, as PEWG did not show statistical significance in relation to effort expectancy, we were unable to support H3.

Next, we examined the HeMo variable, for which our model had an $R^2$ of 0.456 (explaining 45.6% of the variance) and an adjusted $R^2$ of 0.452. According to the guidelines established by Chin [5], it would be described as moderate, and according to Hair et al. [13], it would be described as weak. Regarding the direct effects (noting that the total effects were identical), the statistically significant predictors of the HeMo factor, in order of decreasing strength, were PEWG ($\beta = 0.501$; $p < 0.01$), followed by PIWG ($\beta = 0.247$; $p < 0.01$). Therefore, an increase of 1 unit in PEWG resulted in a 0.501-unit increase in HeMo, consistent with H5. Similarly, each increase of 1 unit in PIWG led to a 0.247-unit increase in HeMo, in line with H6.

Lastly, we analyzed the variable PeEx, which resulted in an $R^2$ of 0.439, explaining 43.9% of the variation, with an adjusted-$R^2$ of 0.435. According to Chin's criteria [5], this would be considered moderate, while Hair et al. [13] would classify it as weak. Regarding direct effects (noting that total effects were identical), the statistically significant antecedents of the PeEx factor, in order of decreasing strength, were PEWG ($\beta = 0.509$; $p < 0.01$), followed by PIWG ($\beta = 0.221$; $p < 0.01$). Therefore, an increase of 1 unit in PEWG led to a 0.509-unit increase in PeEx, consistent with H7. Similarly, each increase of 1 unit in PIWG resulted in a 0.221-unit increase in PeEx, aligning with H8. Table 2 summarizes the results from the structural model and the findings from testing the hypotheses.

**Table 2.** Structural Model Results.

| Path | β (direct) | t-Statistic (direct) | β (total) | t-Statistic (total) | Hypothesis | Decision |
|---|---|---|---|---|---|---|
| Age → BehInt | 0.023 | 0.556 | 0.023 | 0.556 | | |
| Age × EfEx → BehInt | 0.018 | 0.291 | 0.018 | 0.291 | | |
| Age × FC → BehInt | 0.007 | 0.106 | 0.007 | 0.106 | | |
| Age × HeMo → BehInt | 0.083 | 1.143 | 0.083 | 1.143 | | |
| Age × Hab → BehInt | −0.001 | 0.011 | −0.001 | 0.011 | | |

(*continued*)

**Table 2.** (*continued*)

| Path | β (direct) | t-Statistic (direct) | β (total) | t-Statistic (total) | Hypothesis | Decision |
|---|---|---|---|---|---|---|
| Age × PEWG → BehInt | 0.024 | 0.314 | 0.024 | 0.314 | | |
| Age × PIWG → BehInt | −0.054 | 0.746 | −0.054 | 0.746 | | |
| Age × PrVa → BehInt | 0.071 | 1.486 | 0.071 | 1.486 | | |
| Age × SoIn → BehInt | −0.092 | 1.325 | −0.092 | 1.325 | | |
| EfEx → BehInt | −0.039 | 0.374 | −0.039 | 0.374 | | |
| FaCo → BehInt | 0.053 | 0.522 | 0.053 | 0.522 | | |
| Exp → BehInt | 0.059 | 1.149 | 0.059 | 1.149 | | |
| Exp × EfEx → BehInt | −0.114 | 1.133 | −0.114 | 1.133 | | |
| Exp × FC → BehInt | 0.140 | 1.273 | 0.140 | 1.273 | | |
| Exp × HeMo → BehInt | 0.081 | 0.749 | 0.081 | 0.749 | | |
| Exp × Hab → BehInt | −0.018 | 0.256 | −0.018 | 0.256 | | |
| Exp × PEWG → BehInt | −0.139 | 1.417 | −0.139 | 1.417 | | |
| Exp × PIWG → BehInt | 0.013 | 0.143 | 0.013 | 0.143 | | |
| Exp × PV → BehInt | −0.057 | 0.838 | −0.057 | 0.838 | | |
| Exp × SI → BehInt | 0.051 | 0.516 | 0.051 | 0.516 | | |
| Gen → BehInt | 0.142 | 1.759 | 0.142 | 1.759 | | |
| Gen × EfEx → BehInt | 0.073 | 0.480 | 0.073 | 0.480 | | |
| Gen × FC → BehInt | −0.032 | 0.234 | −0.032 | 0.234 | | |
| Gen × HeMo → BehInt | −0.022 | 0.140 | −0.022 | 0.140 | | |
| Gen × Hab → BehInt | 0.034 | 0.360 | 0.034 | 0.360 | | |

(*continued*)

**Table 2.** (*continued*)

| Path | β (direct) | t-Statistic (direct) | β (total) | t-Statistic (total) | Hypothesis | Decision |
|---|---|---|---|---|---|---|
| Gen × PEWG → BehInt | 0.129 | 0.921 | 0.129 | 0.921 | | |
| Gen × PIWG → BehInt | −0.215 | 1.671 | −0.215 | 1.671 | | |
| Gen × PrVa → BehInt | 0.037 | 0.403 | 0.037 | 0.403 | | |
| Gen × SoIn → BehInt | 0.013 | 0.091 | 0.013 | 0.091 | | |
| HeMo → BehInt | 0.329 | 2.663** | 0.329 | 2.663** | | |
| Hab → BehInt | 0.017 | 0.215 | 0.017 | 0.215 | | |
| PeEx → BehInt | 0.302 | 4.271** | 0.302 | 4.271** | | |
| PEWG → BehInt | −0.039 | 0.334 | 0.278 | 2.385* | H1 | Supported[2] |
| PEWG → EfEx | 0.041 | 0.719 | 0.041 | 0.719 | H3 | Not Supported[1] |
| PEWG → HeMo | 0.501 | 7.913** | 0.501 | 7.913** | H5 | Supported[1] |
| PEWG → PeEx | 0.509 | 9.515** | 0.509 | 9.515** | H7 | Supported[1] |
| PEWG × SoIn → BehInt | −0.085 | 1.750 | −0.085 | 1.750 | H9 | Supported[1] |
| PIWG → BehInt | 0.359 | 3.359** | 0.491 | 4.213** | H2 | Supported[1] |
| PIWG → EfEx | 0.418 | 5.383** | 0.418 | 5.383** | H4 | Supported[1] |
| PIWG → HeMo | 0.247 | 3.712** | 0.247 | 3.712** | H6 | Supported[1] |
| PIWG → PeEx | 0.221 | 3.657** | 0.221 | 3.657** | H8 | Supported[1] |
| PIWG × HeMo → BehInt | 0.034 | 0.746 | 0.034 | 0.746 | H10 | Supported[1] |
| PrVa → BehInt | −0.025 | 0.365 | −0.025 | 0.365 | | |
| SoIn → BehInt | 0.056 | 0.483 | 0.056 | 0.483 | | |

*p < 0.05; ** p < 0.01.
[1]supported by direct effects; [2]supported by total effects.

We further investigated whether there are duplicative factors in the original model that could act as substitutes for the introduced warm-glow factors (PEWG and PIWG). To do this, we employed the technique prescribed by Hagedoorn and Wang [12] and utilized moderators. The results indicated that there was no statistically significant moderating role between PEWG and SoIn factors in relation to the dependent variable BehInt. Therefore, H9 was supported. Similarly, no statistically significant role was found between the factors of PIWG and HeMo in relation to the dependent variable BehInt. As a result, H10 was also supported. In essence, we can conclude that the PEWG and PIWG factors are unique within our proposed UTAUT2 + WG model and do not duplicate the factors of social influence and hedonic motivation, respectively.

### 3.3 Explanatory Power, Predictive Ability, and Model Fit for BI

In terms of the dependent variable BehInt, the UTAUT2 + WG model demonstrates significant explanatory power, with an $R^2$ value of 0.706, a metric considered substantial according to both Chin [5] and Hair, Ringle, and Sarstedt [13]. Importantly, comparing the UTAUT2 model with the UTAUT2 + WG model reveals an increase in $R^2$ of 0.049 (from 0.657 to 0.706), and an improvement in adjusted-$R^2$ by 0.042 (from 0.618 to 0.660). This improvement highlights the superior explanatory capability of the UTAUT2 + WG model in understanding the variance in BehInt. Turning to the predictive ability of the UTAUT2 + WG model, the $Q^2$ value for behavioral intention is 0.381. This value exceeding 0 indicates a genuine predictive ability of the latent factors associated with BehInt [16]. Lastly, regarding model fit, the assessment yielded a standardized root mean square residual (SRMR) value of 0.057, comfortably below the acceptable threshold of 0.08, as defined by Hu and Bentler [17]. This outcome strongly suggests that the UTAUT2 + WG model fits the data exceptionally well [14, 15].

## 4 Discussion and Conclusions

This study expands the widely used extended UTAUT2 model by incorporating two constructs—PEWG and PIWG—proposed by Saravanos et al. [26]. These constructs represent the two aspects of warm-glow: EWG, which reflects the satisfaction one receives when being recognized by others for altruistic actions, and IWG, which reflects the satisfaction one experiences when acting altruistically for others. We refer to this enhanced model as UTAUT2 + WG. Our findings confirm that both aspects of warm-glow influence a user's intention to adopt a technology. This supports previous studies that have also examined the positive relationship between the warm-glow phenomenon and technology adoption, albeit from the perspective of older analogous models such as the Theory of Reasoned Action, the Theory of Planned Behavior, and the TAM3.

From the UTAUT2 perspective, the perception of intrinsic warm-glow, as reflected through the PIWG factor, emerges as the most influential determinant of user behavioral intention to adopt the technology. This finding aligns with the observations made by Saravanos et al. [25] in their adaptation of the TAM3 model, TAM3 + WG, where the perception of intrinsic warm-glow similarly held significant sway over user decisions. Thus, supporting the crucial role that this dimension of warm-glow ultimately plays in shaping individual behavior. Similarly, the perception of extrinsic warm-glow, reflected through the PEWG construct, is found to exert a slightly lesser influence on user decisions, ranking fourth within the context of our UTAUT2 + WG model. This finding mirrors the observations of Saravanos et al. [25] and their TAM3 + WG model. Furthermore, our analysis confirms that the PEWG and PIWG factors are distinct and not redundant with comparable factors in the original UTAUT2 model.

Another fascinating finding concerns the concepts of effort expectancy, hedonic motivation, and performance expectancy, which show a mixture of influence from both aspects of warm-glow. It is worth mentioning that our research reveals that IWG, which is closely related to an individual's personal sense of satisfaction and fulfillment, leads users to perceive technology as more user-friendly. It is interesting to note that this effect is not observed in the case of EWG, which is associated with external recognition and validation. This difference may be due to the emphasis placed on outcomes such as gaining recognition or status, rather than on the actual user experience, in situations influenced by EWG.

Hedonic motivation, which represents the pleasure or satisfaction derived from using technology or a service, is influenced by both forms of warm-glow. It is worth noting that EWG generates a larger portion of hedonic motivation compared to the intrinsic dimension of warm-glow. This difference in the impact of EWG and IWG on hedonic motivation highlights the complex interaction between the two categories. The findings indicate that while both aspects of warm-glow contribute to overall hedonic motivation, the external or extrinsic factor appears to have a slightly stronger influence. This nuanced understanding provides valuable insights into their relative significance and impact on user hedonic motivation.

A similar trend arises regarding performance expectancy—the perceived usefulness of the technology. Performance expectancy is positively influenced by both the perception of EWG and IWG. Interestingly, the influence of EWG is slightly more significant in this context as well. Users seem more likely to consider the technology as useful if they can obtain external rewards or recognition from it.

### 4.1 Implications

This work has several implications, starting with its theoretical contributions to the literature on technology adoption. Currently, research on how warm-glow affects technology adoption has mainly relied on generic adoption models like the Theory of Reasoned Action, the Theory of Planned Behavior, and recently, TAM3. Our study advances these concepts by integrating them into the latest adoption model specifically designed for technology, UTAUT2. To the best of our knowledge, this study represents the first empirical application of the PEWG and PIWG constructs to expand the UTAUT2 model for warm-glow. This establishes a foundational framework for researchers interested in using the UTAUT2 model to understand the adoption of technology products that evoke a sense of warm-glow in users, such as the search engine Ecosia.

Our approach integrates both aspects of warm-glow into the UTAUT2 model and introduces an emotional dimension. This provides a deeper insight into how various motivations impact users' decision-making processes and enriches the theoretical understanding of technology adoption. It complements the more traditional rational components of contemporary technology adoption models. Furthermore, it opens up new avenues of research to explore the potential interplay between cognitive and emotional factors. While further exploration and refinement of warm-glow theory can offer valuable insights into the emotional drivers behind technology adoption, our study serves as a foundation for future research to develop the concept of warm-glow within the context of technology adoption.

Additionally, our study serves as a starting point for theoretical exploration into the impact of warm-glow on the perceived usability of technology. We have shown that warm-glow can indeed affect the perceived usability of technology, specifically, how EWG can influence the perceived usefulness derived from its usage, and how both EWG and IWG can impact perceived ease of use and perceived hedonic motivation. Understanding how the dimensions of EWG and IWG influence hedonic motivation and perceived usability enhances our understanding of the emotional factors driving user decisions. This provides a solid foundation for future research in usability. Subsequent studies could examine the interactions between various emotional constructs and how they collectively shape perceptions of usability.

There are also practical implications stemming from this work. Firstly, it demonstrates the feasibility and advantages of designing persuasive technology that incorporates an element of warm-glow. Our findings offer insights into the significant roles played by both the intrinsic and extrinsic aspects of warm-glow, particularly in triggering the intrinsic dimension. This emphasizes the importance of designing technology with a focus on users' emotional needs and desires. Technology developers and designers can utilize this understanding to create persuasive and emotionally resonant experiences that go beyond the mere fulfillment of functional requirements. Tailoring interventions to evoke warm-glow emotions, such as feelings of altruism or personal fulfillment, can enhance user engagement and facilitate long-term adoption. This highlights the potential for technology to positively impact users on an emotional level, thereby fostering deeper connections and sustained usage.

Strategic marketing and communication initiatives that leverage the concept of warm-glow offer promising opportunities. Understanding both the EWG and IWG effects on technology adoption provides valuable insights for guiding marketing and communication efforts. Businesses and organizations can tailor their messaging to emphasize emotional appeals that resonate with diverse user motivations. For example, targeting extrinsic factors like social validation or incentives may attract users driven by external rewards, while highlighting the intrinsic benefits of the technology may appeal to those inclined towards altruism. Furthermore, these insights can also be applied to user training and support. The positive correlation between IWG perceptions and perceived usability suggests that users who see the technology as beneficial to others may also find it more user-friendly. Therefore, developing training and support materials that highlight the societal advantages of the technology can promote a positive user experience and reduce resistance to adoption.

Concluding with a cautionary note, it is important to recognize that warm-glow is a perceptive category [19] and such inherently subjective. Essentially, what evokes warm-glow in one person may not necessarily elicit the same reaction in another; it may even provoke a different emotion. Therefore, careful consideration is necessary to avoid unintentionally alienating certain groups while trying to make a technology offering appealing to others by using warm-glow (e.g., [20]).

### 4.2 Limitations and Future Research Directions

In conclusion, we acknowledge three primary limitations of this study, which simultaneously highlight avenues for further research and development in this field. First, our study's methodology was limited by the selection of a specific solution. We chose an internet search technology that is readily available, free of charge, and devoid of branding. This choice was deliberate to ensure familiarity among survey respondents and to facilitate control over certain dimensions in our analysis. However, future research could further explore the complex relationship between brand perception and warm-glow, the effect of price on warm-glow responses, and the influence of product complexity on warm-glow perceptions. These subtle aspects deserve a more thorough investigation to enhance our understanding of the topic.

The second limitation stems from our sample being exclusively composed of participants from the United States. Given the substantial basis of our research in both the literature on technology acceptance and warm-glow theory, it is essential to recognize the considerable influence of culture in these domains. Cultural factors have been demonstrated to exert a fundamental influence on emotional responses [31], particularly with regard to warm-glow motivations [23]. In fact, the cultural context can significantly shape perceptions of altruism, the assessment of altruistic actions, and the resultant emotional satisfaction experienced by individuals. Moreover, cultural norms and values have a significant impact on attitudes towards technology [8]. Therefore, it is possible that users from different cultural backgrounds may demonstrate different patterns of technology adoption and preferences, influenced by their culturally ingrained attitudes and values. This variation may be seen in distinct behavioral tendencies among technology users in different cultures, as shown in previous research (e.g., [18]). Therefore, it is important to further investigate these cultural nuances to enhance our understanding of this field.

Finally, although our study provided valuable insights, it had an inherently cross-sectional design, which only captured user intentions at a particular moment. It did not investigate whether these intentions translated into actual usage behaviors. To gain a more comprehensive understanding of these dynamics, we recommend conducting a thorough longitudinal examination. This approach would allow us to observe how warm-glow perceptions change over time and how they influence users' intention to adopt technology. It would also shed light on the complex relationship between user intention, usage patterns, and the sustainability of such usage over an extended period. Additionally, exploring concepts such as warm-glow fatigue could enhance our understanding of user engagement dynamics over time.